\documentstyle[11pt]{article}
\newcommand{\vsp}{\vspace*{3mm}}
\newcommand{\be}{\begin{equation}}
\newcommand{\ee}{\end{equation}}
\newcommand{\bd}{\begin{displaymath}}
\newcommand{\ed}{\end{displaymath}}
\newcommand{\bea}{\begin{eqnarray}}
\newcommand{\eea}{\end{eqnarray}}
\newcommand{\sgn}{~{\rm sgn}}

\newcommand{\bra}{\langle}
\newcommand{\ket}{\rangle}

\newcommand{\order}{{\cal O}}
\newcommand{\minus}{\!-\!}
\newcommand{\plus}{\!+\!}
\newcommand{\inner}{\!\cdot\!}
\newcommand{\bnul}{\mbox{\boldmath $0$}}
\newcommand{\bnabla}{\mbox{\boldmath $\nabla$}}
\newcommand{\bm}{\mbox{\boldmath $m$}}
\newcommand{\bq}{\mbox{\boldmath $q$}}
\newcommand{\bx}{\mbox{\boldmath $x$}}
\newcommand{\bA}{\mbox{\boldmath $A$}}
\newcommand{\bB}{\mbox{\boldmath $B$}}

\newcommand{\bD}{\mbox{\boldmath $D$}}
\newcommand{\bF}{\mbox{\boldmath $F$}}
\newcommand{\bG}{\mbox{\boldmath $G$}}
\newcommand{\bK}{\mbox{\boldmath $K$}}
\newcommand{\bI}{\mbox{\boldmath $I$}}
\newcommand{\bJ}{\mbox{\boldmath $J$}}
\newcommand{\bL}{\mbox{\boldmath $L$}}

\newcommand{\bR}{\mbox{\boldmath $R$}}

\newcommand{\bxi}{\mbox{\boldmath $\xi$}}
\newcommand{\bXi}{\mbox{\boldmath $\Xi$}}
\newcommand{\bsigma}{\mbox{\boldmath $\sigma$}}

\begin{document}
\title{\bf Finite Size Effects in \\
Separable Recurrent Neural Networks}
\author{A. Castellanos$^\dag$, A.C.C. Coolen$^\ddag$ and L. Viana$^\S$}
\maketitle

\begin{center}
\dag~CICESE, F\'\i sica de Materiales \\
A. Postal 2681, Ensenada 22800, B.C. , M\'exico \\
and \\
Dept. de  F\'\i sica, Universidad de Sonora \\
A. Postal 1626, Hermosillo 83000, Son., M\'exico.
\end{center}
\begin{center}
\ddag~ Dept. of Mathematics, King's College, University of London\\
Strand, London WC2R 2LS, U.K.
\end{center}
\begin{center}
\S ~CECIMAT, UNAM\\
A. Postal 2681, 22800 Ensenada, B.C., M\'{e}xico
\end{center}

\vsp
\begin{center}
PACS: 87.30, 05.20
\end{center}
\vsp

\begin{abstract}
We perform a systematic analytical study of finite size effects
in separable recurrent neural network models with sequential
dynamics, away from saturation.  We find two types of finite size
effects: thermal fluctuations, and disorder-induced `frozen'
corrections to the mean-field laws.  The finite size effects  are described by
equations that correspond to a time-dependent
Ornstein-Uhlenbeck process.  We show how the theory can be used to
understand and quantify various finite size phenomena in recurrent neural
networks, with and without detailed balance.
\end{abstract}

\pagebreak\tableofcontents

\pagebreak\section{Introduction}

Infinite-range spin models of recurrent neural networks, with
information stored in
the values of the interaction strengths of pairs of spins,
have been studied intensively with statistical mechanical tools
following the papers  \cite{hopfield} and
\cite{amitetal}. The first wave of such studies involved mainly
equilibrium analyses, and was consequently restricted to models obeying
detailed balance. Away from the saturation regime
(where small numbers of patterns are stored) such models can be
solved with standard mean-field techniques and display standard mean-field
behaviour. In contrast, in the saturation regime (where an extensive
number of patterns are stored) tools from spin-glass theory are
required (replica theory), and non-trivial phases occur.
The second wave of studies employed tools from non-equilibrium
statistical mechanics. Here restriction to detailed balance models is
irrelevant. However, in view of the highly non-trivial nature of
the glassy non-ergodic dynamics of models close to saturation, most dynamical
studies have been restricted to recurrent neural networks with only
small
numbers of patterns stored. For an overview of the relevant literature
see textbooks such as \cite{physicsofnn1} or reviews such as
\cite{reviewaccds}.

In spite of the fact that finite size effects have been reported
regularly in literature, and that they are know to persist even for
system sizes up to $N\approx 10^5$ \cite{Kohring}, it appears that
systematic studies of finite size effects in recurrent networks
which go beyond pilot studies such as \cite{chicago} have not yet
been performed. The purpose of this paper is to carry out a
comprehensive analysis of finite size effects (in first non-trivial
order in the system size) for a reasonably general class of
recurrent neural network models, where the interaction matrix has
a separable structure. This class contains detailed balance models,
as well as
models without detailed balance.
Away from saturation, finite size effects in these systems take
the form of mainly 
thermal fluctuations of order $O(N^{-1/2})$ around mean field
trajectories for dynamical order parameters,
as well as disorder-induced `frozen'
corrections to the mean-field laws.
Close to saturation
even the $N\to\infty$ dynamics cannot be solved in explicit form
(describing transients is found to necessitate approximations in all of the
present approaches; in the path integral formalism
\cite{horneretal,riegeretal} as well as
in dynamical replica theory \cite{drt}). Therefore in the latter
regime the development of a finite size theory would be premature.

We study the evolution of finite recurrent neural
network models
away from saturation and with Glauber-type (stochastic) neuronal  dynamics.
We expand the Kramers-Moyal
expansion for the system's natural dynamic order parameters on finite
time-scales, and calculate the statistical properties of
finite size effects to first non trivial order in $1/N$. The finite size effects
 turn out to
be governed by a time-dependent Ornstein-Uhlenbeck process.
Our theory is used to analyse  the
dependence of finite size effects on detailed balance,
scaling properties of fluctuations close to phase boundaries, and
escape processes in critical models which are driven purely by finite
size effects.
Comparison with extensive numerical simulations confirms the
theoretical predictions in all cases.

\section{Derivation of Macroscopic Laws}

\subsection{Model Definitions and Simple Relations}

We consider a system composed of a large, but finite, number $N$
of interconnected neurons, modeled as Ising spins $\sigma_i\in\{-1,1\}$.
The vector $\bsigma (t)=(\sigma_1(t), \dots , \sigma_N(t))$
defines the state of the system at time $t$.
The dynamics of the system is defined
by a master equation for the microscopic probability distribution
$p_t(\bsigma)$:
\be
\frac{d}{dt}p_t(\bsigma) =
\sum_i\left\{
w_i(F_i\bsigma)p_t(F_i\bsigma)\minus
w_i(\bsigma)p_t(\bsigma)\right\}
\label{eq:mastereq}
\ee
\be
w_i(\bsigma)=\frac{1}{2}[1\minus\sigma_i\tanh(\beta h_i(\bsigma))],
\label{eq:rates}
\ee
where $w_i(t)$ defines the rate of the single-spin transitions
$\sigma_i(t) \to - \sigma_i(t)$, $\beta=T^{-1}$ (the inverse
temperature) controls  the stochasticity in the dynamics,
and $F_i$  is an operator that flips the $i-$th spin, i.e.
$F_i f(\sigma_1,\ldots,\sigma_N)=
f(\sigma_1,\ldots,\minus\sigma_i,\ldots,\sigma_N)$.
The local field
$h_i(\bsigma)$ is given by the usual linear expression
\be
h_i(\bsigma)=\sum_j J_{ij}\sigma_j \plus \theta_i
\label{eq:extfield}
\ee
where $J_{ij}$ is the strength of the synaptic connection from
neuron $j$ to neuron $i$, and $\theta _i$ is a response threshold.
The interactions $J_{ij}$ are assumed to result from a learning process involving
a finite number $p$ 
of randomly chosen binary patterns
$\bxi^\mu=(\xi_1^\mu,\ldots,\xi_N^\mu)\in\{-1,1\}^N$,
with $\mu = 1,\dots , p$. We restrict ourselves to situations
where the interactions have a separable form (see e.g.
\cite{separable1,separable2,separable3})
and  introduce a parameter  $\Delta\in\{0,1\}$ to control whether or not
self-interactions $J_{ii}$  will be allowed:
\be
J_{ij}=\left[1\minus \Delta \delta_{ij}\right]\frac{1}{N}
\sum_{\mu\nu=1}^p \xi_i^\mu A_{\mu\nu}
\xi_j^\nu~~~~~~~~~~~~
\left\{
\begin{array}{ll} \Delta=0: & J_{ii}\neq 0\\ \Delta=1: &
J_{ii}=0\end{array}
\right.
\label{eq:interactions}
\ee

Given the process (\ref{eq:mastereq},\ref{eq:rates}), we can define
averages over the microscopic ensemble in the usual way, and find simple
relations for the temporal derivatives of such averages:
\be
\bra f(\bsigma)\ket_t  \equiv  \sum_{\bsigma} p_t(\bsigma)f(\bsigma)
~~~~~~~~
\frac{d}{dt}\bra f(\bsigma)\ket_t
=\bra \sum_i w_i(\bsigma)
\left[f(F_i\bsigma)\minus f(\bsigma)\right]\ket_t
\label{eq:ddtaverage}
\ee
In particular, application to $f(\bsigma)=\sigma_k$ gives
\be
\frac{d}{dt}\bra \sigma_k\ket_t=\bra \tanh(\beta
h_k(\bsigma))\ket_t-\bra\sigma_k\ket_t
\label{eq:ddtsigmak}
\ee
In order to make the transition to a macroscopic description of the
process, we define the usual  pattern overlaps. These observables
(which for finite $N$ take discrete values only) measure the similarity
between the
state of the system and each of the $p$ stored patterns:
\be
\bm (\bsigma ) = (m_1 (\bsigma ), \dots , m_p (\bsigma) )
~~~~~~~~~~~~~
m_\mu(\bsigma) = \frac{1}{N} \sum_i^N \xi^\mu_i \sigma_i
\label{eq:overlaps}
\ee
The probability density for the macroscopic variables $\bm$ is given
by:
\be
P_t(\bm) \equiv \sum_{\bsigma}
p_t(\bsigma)\delta\left[\bm\minus \bm(\bsigma)\right]
\label{eq:macrodistribution}
\ee
We next define conditional, or sub-shell, averages of observables
$f(\bsigma)$. These are averages over the statistical ensemble,
with the microscopic probability distribution $p_t(\bsigma)$,
restricted to those micro-states $\bsigma$ that obey
$\bm(\bsigma)=\bm$ (in a distributional sense):
\be
\bra f(\bsigma)\ket_{\bm;t} \equiv \frac{
\sum_{\bsigma}p_t(\bsigma)\delta[\bm\minus \bm(\bsigma)] f(\bsigma)}{
\sum_{\bsigma}p_t(\bsigma)\delta[\bm\minus \bm(\bsigma)]}
\label{eq:subshellaverage}
\ee
Note that the $\delta$-distribution in the
definition (\ref{eq:subshellaverage}) allows us to replace
all occurrences of $\bm(\bsigma)$ in $f$ simply by $\bm$:
\be
\bra f[\bsigma,\bm(\bsigma)]\ket_{\bm;t}=\bra
f[\bsigma,\bm]\ket_{\bm;t}
\label{eq:identity1}
\ee

\subsection{The Kramers-Moyal Expansion}

The dynamic equation for the macroscopic variables $\bm (\bsigma)$
can be obtained by making the choice $f(\bsigma)=\delta[\bm-\bm(\sigma)]$
in equation
(\ref{eq:ddtaverage})
\bd
\frac{d}{dt}P_t(\bm)=\sum_{\bsigma}p_t(\bsigma)\sum_i w_i(\bsigma)
\left\{\delta[\bm\minus \bm(\bsigma)\plus \frac{2}{N}\sigma_i\bxi_i]
\minus \delta[\bm\minus \bm(\bsigma)]
\right\}
\ed
Inside this expression we make a Taylor expansion
in powers of the
vector $\frac{2}{N}\sigma_i\bxi_i$ and write the result
in terms of sub-shell averages (\ref{eq:subshellaverage}):
\be
\frac{d}{dt}P_t(\bm)=
\sum_{\ell\geq 1} \frac{1}{\ell
!}\sum_{\mu_1=1}^p\cdots\sum_{\mu_\ell=1}^p
\frac{\partial^\ell}{\partial m_{\mu_1} \cdots\partial m_{\mu_\ell}}
\left\{ P_t(\bm) F^{(\ell)}_{\mu_1,\ldots,\mu_\ell}[\bm;t]
\right\}
\label{eq:kramersmoyal}
\ee
\be
F^{(\ell)}_{\mu_1,\ldots,\mu_\ell}[\bm;t]=
\bra \sum_i w_i(\bsigma)
\left[\frac{2}{N}\sigma_i\xi^{\mu_1}_i\right]
\cdots
\left[\frac{2}{N}\sigma_i\xi^{\mu_\ell}_i\right]\ket_{\bm;t}=
\order(N^{1-\ell})
\label{eq:termsinKM}
\ee
This is the so-called Kramers-Moyal expansion, applied to the present class of
models.

Since equation (\ref{eq:kramersmoyal}) cannot be solved exactly,
we follow the standard procedure for `large systems' (see e.g.
\cite{vankampen}), and expand (\ref{eq:kramersmoyal})
in powers of $1/N$. By keeping only the two leading orders,
we obtain, at least on finite time-scales  (i.e. on times not scaling
with the system size $N$),
the $N\rightarrow\infty$ (mean-field) equations plus the
leading order contribution  due to the finite size effects
\footnote{Note that due to Pawula's Theorem \cite{pawula} we have but
three options: 1) to retain only the $\order(N^0)$ terms of (\ref{eq:kramersmoyal}),
(describing the infinite system), 2) to include in addition the $\order(N^{-1})$ terms,
or 3) to keep all
remaining orders in $N$. Retaining a finite number of terms, until and including
order $N^{-n}$ with $n>1$ would generate solutions $P_t (\bm)$
which violate the obvious condition that they be positive
definite (i.e. represent probability densities).}:
\be
\frac{d}{dt}P_t(\bm)=
\sum_{\mu=1}^p
\frac{\partial}{\partial m_{\mu}}
\left\{ P_t(\bm) F^{(1)}_{\mu}[\bm;t]
\right\}+\frac{1}{2}\sum_{\mu\nu=1}^p
\frac{\partial^2}{\partial m_{\mu}\partial m_{\nu}}
\left\{ P_t(\bm) F^{(2)}_{\mu\nu}[\bm;t]
\right\}+\order(N^{-2})
\label{eq:truncatedkramersmoyal}
\ee
Upon insertion
of the transition rates $w_i(\bsigma)$ (\ref{eq:rates}) and the
local fields (\ref{eq:extfield}), written in terms of the overlaps
(\ref{eq:overlaps}), we can work
out explicitly and simplify (with e.g. (\ref{eq:identity1}))  the
various terms, giving
\bd
F^{(1)}_\mu[\bm;t]
=m_\mu \minus
\frac{1}{N}\sum_i \xi^{\mu}_i
\tanh\beta\left[\bxi_i \inner\bA \bm
\plus\theta_i\right] ~~~~~~~~~~~~~~
\ed
\be
+ \frac{\beta\Delta}{N^2}\sum_i \bra \sigma_i\ket_{\bm;t} \,
\xi^{\mu}_i (\bxi_i\inner
\bA\bxi_i) \left[1\minus \tanh^2\beta\left[\bxi_i \inner\bA \bm
\plus\theta_i\right]\right]
\plus\order(N^{-2})
\label{eq:flowtermexpanded}
\ee
Similarly:
\be
\frac{1}{2}F^{(2)}_{\mu\nu}[\bm;t]
=  \frac{1}{N^2}\sum_i\xi^{\mu}_i\xi^{\nu}_i
\left[1\minus\bra\sigma_i\ket_{\bm;t}\tanh\beta(
\bxi_i \inner\bA \bm
\plus \theta_i)\right]+\order(N^{-2})
\label{eq:diffusiontermexpanded}
\ee
In the limit $N\rightarrow\infty$  equation 
(\ref{eq:truncatedkramersmoyal}) 
reduces to a Liouville equation:
\bd
\frac{d}{dt}P_t(\bm)=
\sum_{\mu=1}^p
\frac{\partial}{\partial m_{\mu}}
\left\{ P_t(\bm) \left[m_\mu \minus \bra \xi_\mu
\tanh\beta\left[\bxi\inner\bA \bm
\plus\theta\right]\ket_{\bxi,\theta}\right]
\right\}
\ed
with the deterministic solution
\be
P_t(\bm)=\delta\left[\bm-\bm^\ast (t)\right]~~~~~~~~
\frac{d}{dt}\bm^\ast  (t)=
\bra \bxi
\tanh\beta\left[\bxi\inner\bA \bm^\ast (t)
\plus\theta\right]\ket_{\bxi,\theta}-\bm^\ast (t)
\label{eq:deterministic}
\ee
where we defined
\bd
\bra g[\bxi,\theta]\ket_{\bxi,\theta} =
\lim_{N\rightarrow\infty} \frac{1}{N}\sum_i g[\bxi_i,\theta_i]
~~~~~~~\bxi=(\xi_1,\dots,\xi_p)
\ed
with $\bxi_i=(\xi_i^1,\ldots,\xi_i^p)$.

\section{Description via Rescaled Variables}

\subsection{Derivation of Fokker-Planck Equation}

The stochastic vector $\bm(\bsigma)$ can apparently be
written as the sum of a deterministic term $\bm^\ast (t)$ and a
fluctuating term, with the latter vanishing for
$N\rightarrow\infty$. Since for
mean-field models the
overlaps can be seen as an average over $N$ independent random variables,
one would expect,
from the central limit theorem, the fluctuating term to scale as
$N^{-\frac{1}{2}}$. Therefore we define a  new
stochastic variable $\bq(\bsigma)$:
\be
\bq(\bsigma)=\sqrt{N}\left[\bm(\bsigma)\minus\bm^\ast (t)\right]~~~~~~~~
{\cal P}_t(\bq)=\int\! d\bm~P_t(\bm)\delta\left[\bq- \sqrt{N}[\bm\minus \bm^\ast(t)]\right]
\label{eq:newvariables}
\ee
in which $\bm^\ast (t)$ is the solution of the deterministic equation
(\ref{eq:deterministic}). Working out the temporal derivative of ${\cal
P}_t(\bq)$, with the help of the macroscopic equation
(\ref{eq:truncatedkramersmoyal}),  and
taking the  $N \to \infty$ limit, leads to a convenient description
of the leading $\order(N^{-\frac{1}{2}})$ finite size effects
in terms of a Fokker-Planck
equation for the rescaled variables $\bq$:
\bd
\bm(\bsigma(t))
=\bm^\ast (t)+\frac{1}{\sqrt{N}}\bq(t)+\order(\frac{1}{N})
\ed
\be
\frac{d}{dt}{\cal P}_t(\bq)=
\sum_{\mu} \frac{\partial}{\partial q_{\mu}}\left\{
{\cal P}_t(\bq) F_\mu[\bq;t]\right\}
+\sum_{\mu\nu}\frac{\partial^2}{\partial q_{\mu}\partial q_{\nu}}\left\{
{\cal P}_t(\bq) D_{\mu\nu}[\bq;t]\right\}
\label{eq:fokkerplanck}
\ee
in which the flow term is given by
\bd
F_\mu[\bq;t]=
q_\mu\minus \beta
\bra\xi_{\mu}(\bxi\inner\bA \bq)
\left[1\minus\tanh^2\beta[\bxi\inner\bA\bm^\ast
(t)\plus\theta]\right]\ket_{\bxi,\theta}
~~~~~~~~~~~~~~~~~
\ed
\bd
~~~~~~~~~~~~~~~~~
+ \lim_{N\rightarrow\infty}\sqrt{N}\left\{
 \bra \xi_{\mu}
\tanh\beta[\bxi \inner\bA \bm^\ast (t)\plus\theta]\ket_{\bxi,\theta}
\minus
\frac{1}{N}\sum_i\xi^{\mu}_i\tanh\beta\left[\bxi_i\inner\bA \bm^\ast (t)\plus\theta_i\right]
\right\}
\ed
where we used (\ref{eq:flowtermexpanded}),  (\ref{eq:deterministic})
and (\ref{eq:newvariables}).
The last term in $F_\mu[\bq;t]$ describes a `frozen' finite size
correction to the flow field, depending on the microscopic
realization of the pattern components.
Similarly, we can work out the diffusion matrix:
\be
D_{\mu\nu}[\bq;t]
 =  \lim_{N\rightarrow\infty}\frac{1}{N}\sum_i\xi^{\mu}_i\xi^{\nu}_i
\left[1\minus\bra\sigma_i\ket_{\bm^\ast (t);t}\tanh\beta[
\bxi_i \inner\bA \bm^\ast (t)
\plus \theta_i]\right]
\ee
where we have used
$\bra\sigma_i\ket_{\bm , t}= \bra\sigma_i\ket_{\bm^\ast
(t);t}+\order(N^{-\frac{1}{2}})$.
According to (\ref{eq:ddtsigmak}) we may also use
\be
\bra\sigma_i\ket_{\bm^\ast (t);t}=
\sigma_i(0)e^{-t}+\int_0^t\!ds~
e^{s-t}\tanh\beta[\bxi_i\inner\bA\bm^\ast (s)\plus\theta_i]+\order(\frac{1}{\sqrt{N}})
\label{eq:microspins}
\ee
We conclude that the diffusion matrix does not depend on $\bq$:
\bd
D_{\mu\nu}(t)
=  \bra\xi_\mu \xi_\nu\ket_{\bxi,\theta}
-e^{-t}\lim_{N\rightarrow\infty}\frac{1}{N}\sum_i
\xi^{\mu}_i\xi^{\nu}_i
\sigma_i(0)
\tanh\beta[\bxi_i\inner\bA \bm^\ast (t)\plus \theta_i]
~~~~~~~~~
\ed
\be
~~~~~~~~~
-  \int_0^t\!ds~e^{s-t}\bra \xi_{\mu}\xi_{\nu}
\tanh\beta[\bxi\inner\bA\bm^\ast (s)\plus\theta]
\tanh\beta[\bxi\inner\bA \bm^\ast (t)\plus\theta]\ket_{\bxi,\theta}
\label{eq:Dmunu}
\ee
The fact that the above limits exist and give a non-trivial
flow-term and diffusion matrix for (\ref{eq:fokkerplanck}) is the
a-posteriori justification of the ansatz  (\ref{eq:newvariables}).
\vsp

We finally have to specify the pattern- and threshold statistics
in order to analyze the Fokker-Planck
equation (\ref{eq:fokkerplanck}). We choose independently drawn
pattern components $\xi_i^\mu\in\{\minus 1,1\}$ (with equal
probabilities) and independently drawn neural thresholds $\theta_i$ (with
probability distribution $W(\theta)$):
\be
\bra g[\bxi,\theta]\ket_{\bxi,\theta}=
2^{-p}\!\!\!\sum_{\bxi\in\{-1,1\}^p}\!\!\int\!d\theta~W(\theta)g[\bxi,\theta]
\label{eq:randompatterns}
\ee
which gives convenient relations like
\bd
\bra f(\bxi) g(\theta)\ket_{\bxi,\theta}=
2^{-p}\!\!\sum_{\bxi\in\{-1,1\}^p}\!\!
f(\bxi)~ \int\!d\theta~W(\theta)g(\theta)
~~~~~~~~~
\bra \xi_\mu\ket_{\bxi,\theta}=0
~~~~~~~~~
\bra \xi_\mu\xi_\nu \ket_{\bxi,\theta}=\delta_{\mu\nu}
\ed
The flow term $\bF[\bq;t]$ in (\ref{eq:fokkerplanck}) can be written as the sum
of two contributions; the first ($\bK$) depends  on the specific microscopic
realization of the pattern components, whereas the second
(with the matrix $\bL$) depends
only on the probability distribution of the pattern components:
\be
F_\mu[\bq;t]=K_\mu(t) +\sum_\nu L_{\mu\nu}(t) q_\nu
\label{eq:flowtermrandom}
\ee
\be
K_\mu(t)= \lim_{N\rightarrow\infty}\sqrt{N}\left\{
 \bra \xi_{\mu}
\tanh\beta[\bxi \inner\bA\bm^\ast(t)\plus\theta]\ket_{\bxi,\theta}
\minus
\frac{1}{N}\sum_i\xi^{\mu}_i\tanh\beta\left[\bxi_i\inner\bA\bm^\ast(t)
\plus\theta_i\right]
\right\}
\label{eq:Kmu}
\ee
\be
L_{\mu\nu}(t) =  \delta_{\mu\nu}- \beta \sum_{\lambda}
\bra\xi_{\mu}\xi_\lambda
\left[1\minus\tanh^2\beta[\bxi\inner\bA\bm^\ast(t)\plus\theta]\right]\ket_{\bxi,\theta}
A_{\lambda\nu}
~~~~~~~~~~~~~~~~~~~~~~~~~~~~~~~
\label{eq:Lmunu}
\ee
The diffusion matrix in (\ref{eq:fokkerplanck}) is symmetric, and
can be simplified to
\bd
D_{\mu\nu}(t)
=  \delta_{\mu\nu}- e^{-t}
\lim_{N\rightarrow\infty}\frac{1}{N}\sum_i\xi^{\mu}_i\xi^{\nu}_i
\sigma_i(0)\tanh\beta[\bxi_i\inner\bA \bm^\ast(t)\plus\theta_i]
~~~~~~~
\ed
\be
~~~~~~~ - \int_0^t\!ds~e^{s-t}\bra\xi_{\mu}\xi_{\nu}
\tanh\beta[\bxi\inner\bA\bm^\ast(s)\plus\theta]
\tanh\beta[\bxi\inner\bA\bm^\ast(t)\plus\theta]\ket_{\bxi,\theta}
\label{eq:diffusionmatrixrandom}
\ee
Equation (\ref{eq:fokkerplanck}),
with its flow term (\ref{eq:flowtermrandom}) which is linear in the
rescaled fluctuation variables
$\bq$ and with its $\bq$-independent diffusion matrix
(\ref{eq:diffusionmatrixrandom}),
is called the `linear noise' Fokker-Planck  equation;
it describes a so-called time dependent Ornstein-Uhlenbeck
process (see e.g. \cite{gardiner}).

\subsection{General Solution}

The natural solution of the Ornstein-Uhlenbeck process
(\ref{eq:fokkerplanck}) is a Gaussian distribution:
\be
{\cal P}_t(\bq)= \frac{1}{(2 \pi )^{p/2} \sqrt{ \det  \bXi(t)}}~
\exp \left\{ - \frac{1}{2}[\bq -  \bra \bq \ket_t]\cdot
\bXi^{-1}(t)[\bq -
\bra \bq\ket_t] \right\}
\label{eq:qdistribution}
\ee
It is fully characterized in the usual way by the time-dependent average
$\bra \bq\ket_t$ and the time-dependent correlation matrix
\be
\Xi_{\mu \nu} (t) = \bra q_\mu q_\nu \ket_t -
\bra q_\mu \ket_t  \bra q_\nu \ket_t
\label{eq:central}
\ee
Here we denote averages over the distribution
(\ref{eq:qdistribution}) as  $\bra f(\bq)\ket_t=\int\!d\bq~{\cal
P}_t(\bq)f(\bq)$. Insertion of equation (\ref{eq:qdistribution}) as an
ansatz into the Fokker-Planck equation (\ref{eq:fokkerplanck}) gives
the following three necessary and sufficient conditions for
(\ref{eq:qdistribution}) to be a solution:
\be
\frac{d}{dt} \bra \bq\ket_t = -\bL(t) \bra\bq\ket_t-\bK(t)
\label{eq:ddtq}
\ee
\be
\frac{d}{dt}\bXi(t)= -\bL(t)\bXi(t)-\bXi(t)\bL^\dag(t)+2\bD(t)
\label{eq:ddtqq}
\ee
\be
\frac{d}{dt}\log\det\bXi(t)+2~{\rm Tr}[\bL\minus \bD\bXi^{-1}]=\bnul
\label{eq:thirdcondition}
\ee
with the (symmetric) diffusion matrix
(\ref{eq:diffusionmatrixrandom}).
Equations (\ref{eq:ddtq},\ref{eq:ddtqq}) define the  evolution in time of the
moments of the distribution (\ref{eq:qdistribution}). Equation
(\ref{eq:thirdcondition}) is then solved automatically, which can be
seen by combining the Wronski identity $\frac{d}{dt}\log\det\bB ={\rm
Tr}[\bB^{-1}\frac{d}{dt}\bB]$ with equation (\ref{eq:ddtqq}).
\vsp

Since the differential equations (\ref{eq:ddtq},\ref{eq:ddtqq})
are linear, they can be solved using standard procedures (see e.g.
\cite{vankampen,gardiner}). One defines the
propagator $\bG(t)$ as the (matrix) solution of
\be
\frac{d}{dt}\bG(t)=-\bL(t)\bG(t),~~~~\bG(0)=\bI
~~~~~~~~
{\rm or}
~~~~~~~~
\bG (t) = \bI  - \int_0^t\!ds~ \bL (s) \bG (s)
\label{eq:propagator}
\ee
in which $\bI$ denotes the unit matrix.
This allows us to express the solution of
(\ref{eq:ddtq},\ref{eq:ddtqq}) in the following compact way:
\be
\bra \bq\ket_t=\bG(t)\bra\bq\ket_0-\bG(t)\int_0^t\!ds~\bG^{-1}(s)\bK(s)
\label{eq:qsolution}
\ee
\be
\bXi(t)=\bG(t)\bXi(0)\bG^\dag(t)+
2\bG(t)\int_0^t\!ds~\bG^{-1}(s)\bD(s)[\bG^\dag(s)]^{-1}\bG^\dag(t)
\label{eq:qqsolution}
\ee
(as can be verified by insertion).
Note, however, that calculating the propagator
(\ref{eq:propagator}) can still be non-trivial.

\subsection{Stationary States and Detailed Balance}

For large times $t\rightarrow\infty$ the dependence of
(\ref{eq:diffusionmatrixrandom}) on the
microscopic initial
conditions vanishes. Furthermore, for macroscopic stationary states,
i.e. $\bm^\ast (t)=\bm^\ast$ for all $t$, with $\bm^\ast$
given by the solution of the macroscopic fixed-point equation
\be
\bm^\ast =\bra\bxi\tanh\beta[\bxi\cdot\bA\bm^\ast
\plus\theta]\ket_{\bxi,\theta}
\label{eq:macrofp}
\ee
we can in addition
perform the time
integration in (\ref{eq:diffusionmatrixrandom}). All flow- and
diffusion terms in the Fokker-Planck equation (\ref{eq:fokkerplanck})
become independent of time, the convection matrix (\ref{eq:Lmunu})
can be expressed in terms of the diffusion matrix as
\be
\bL=\bI-\beta \bD\bA
\label{eq:stationaryL}
\ee
and our process (\ref{eq:fokkerplanck}) reduces to a time
independent Ornstein-Uhlenbeck process, characterised by
\bd
\bF[\bq]=\bK\plus[\bI \minus \beta \bD\bA]\bq~~~~~~~~~~~~
D_{\mu\nu}
= \delta_{\mu\nu}
\minus \bra\xi_{\mu}\xi_{\nu}
\tanh^2\beta[\bxi\inner\bA\bm^\ast\plus\theta]\ket_{\bxi,\theta}
\ed
\bd
\bK=\lim_{N\rightarrow\infty}\sqrt{N}\left\{
 \bra \bxi
\tanh\beta[\bxi \inner\bA\bm^\ast \plus\theta]\ket_{\bxi,\theta}
\minus
\frac{1}{N}\sum_i\bxi_i\tanh\beta\left[\bxi_i\inner\bA\bm^\ast \plus\theta_i\right]
\right\}
\ed
Since the matrix $\bL$ (\ref{eq:stationaryL}) is stationary, the
propagator (\ref{eq:propagator}) reduces to $\bG(t)=\exp[-t\bL]$.
Whether or not a macroscopic stationary state $\bm^\ast(t)=\bm^\ast$
will be reached
will depend on the
choice made for the matrix $\bA$.
\vsp

A sufficient condition for asymptotic stationarity is
(microscopic) detailed balance, which
states that, in addition to stationarity of the probability
distribution $p_t(\bsigma)$, there is no net probability current
between any two configurations  $\bsigma$ and $\bsigma^\prime$.
For the models studied in the present  paper this translates into
symmetry of the matrix $\bA$ and absence of self-interactions,
i.e. $\Delta=1$ (apart from pathological exceptions, like systems with
self-interactions only). See e.g.
\cite{laughtoncoolen,lecturenotes}. Note, however, that our equations
(\ref{eq:deterministic},\ref{eq:fokkerplanck})
show that presence
or absence of self-interactions does not yet play a role in the first
two leading orders in the system size.
We will now inspect the conditions for the Fokker-Planck
equation (\ref{eq:fokkerplanck}) to have a stationary solution, and
show that for this solution to obey detailed balance
(i.e. for there to be 
no net probability current in $\bq$-space) we must again require
symmetry of the matrix $\bA$.
Equation (\ref{eq:fokkerplanck}) can be written as a continuity equation for
the probability density $P_t(\bq)$:
\bd
\frac{d}{dt}P_t(\bq)+\bnabla\cdot\bJ_t(\bq)=0
\ed
with $\bnabla=(\frac{\partial}{\partial
q_1},\ldots,\frac{\partial}{\partial q_p})$ and with,
in the case of macroscopic stationarity,
\be
\bJ_t(\bq)=P_t(\bq)\left\{\bD\bXi^{-1}(\bq-\bra\bq\ket_t)-\bK-\bL\bq\right\}
\label{eq:current}
\ee
>From (\ref{eq:ddtq},\ref{eq:ddtqq}) we deduce that for
(\ref{eq:qsolution}) to be a stationary solution, i.e.
$\frac{d}{dt}\bra\bq\ket=\bnul$ and $\frac{d}{dt}\bXi=\bnul$, we must require
\be
\bra\bq\ket=-\bL^{-1}\bK~~~~~~~~~~~~~~
\frac{1}{2}\left\{\bL\bXi +(\bL\bXi)^\dag\right\}=\bD
\label{eq:stationarityofP}
\ee
For such stationary states the probability current (\ref{eq:current})
reduces to
\be
\bJ(\bq)=P(\bq)~[\bD\bXi^{-1}\minus \bL]
(\bq\plus \bL^{-1}\bK)
\label{eq:stationarycurrent}
\ee
We conclude that detailed balance, i.e. a vanishing current, requires
in addition to (\ref{eq:stationarityofP}) that
\be
\bL\bXi=\bD
\label{eq:stronger}
\ee
Combination of
(\ref{eq:stationarityofP},\ref{eq:stronger}) leads to the condition
$\bD\bL^\dag=\bL\bD$, which, with identity
(\ref{eq:stationaryL}), translates into
$\bD\bA^\dag\bD= \bD\bA\bD$.
We now use the symmetry and non-negativity of the
stationary diffusion matrix $\bD$, i.e. $\bx\cdot\bD\bx=\bra
(\bx\cdot\bxi)^2[1\minus
\tanh^2\beta[\bxi\inner\bA\bm^\ast\plus\theta]]\ket_{\bxi,\theta}\geq
0$ (with $\bx\cdot\bD\bx=0$ only for $\bx=\bnul$). We denote with
$\{|n\ket\}$ the orthogonal basis of normalised eigenvectors of $\bD$,
and with $\{d_n\}$ the corresponding (positive) eigenvalues. This allows
us to  derive
from $\bD\bA^\dag\bD= \bD\bA\bD$ that $\forall n,m:~d_nd_m
\bra n|[\bA^\dag\minus \bA]|m\ket=0$. This implies that $\bA=\bA^\dag$, which thus is
found to be not  only a
sufficient condition, but also a necessary
condition for a stationary solution of equation
(\ref{eq:fokkerplanck}) to obey detailed balance.

\section{Application to Associative Memories}

Our first application is an associative memory model, which
generalizes the standard model of \cite{hopfield} by allowing for
patterns to be stored with different embedding strengths:
$J_{ij}=\frac{1}{N}[1\minus\Delta\delta_{ij}]\sum_{\mu}w_\mu \xi_i^\mu\xi_j^\mu$ (with $0<w_\mu\leq 1$ for all
$\mu$). This model, due to
\cite{LV}, corresponds to the choice $A_{\mu\nu}=w_\mu \delta_{\mu\nu}$
in the language of (\ref{eq:interactions}), and thus obeys
detailed balance. For simplicity we choose zero thresholds, i.e.
$W(\theta)=\delta(\theta)$. We will only study finite
size corrections to the so-called `pure states', where
$m^\ast(t)=m(t)\delta_{\lambda\mu}$, which are the most important
macroscopic solutions from both a thermodynamic and an
information processing point of
view. Without loss of generality we can choose $m(0)>0$ and $\lambda=1$
(as long as
we refrain from ordering the embedding strengths with respect to magnitude), so
\be
m^\ast_\mu(t)=m^\ast(t)\delta_{\mu,1}
~~~~~~~~~
\frac{d}{dt} m^\ast(t)=\tanh\beta[ w_1 m^\ast(t)]-m^\ast(t)
\label{eq:macrolawLV}
\ee
The above mean-field equation (\ref{eq:macrolawLV}) 
will always evolve towards a
fixed-point, given by the solution of $m^\ast=\tanh\beta[
w_1 m^\ast]$. Above the critical temperature $T_c=w_1$
the macroscopic fixed-point is paramagnetic, i.e. $m^\ast=0$,
below $T_c$ one finds an ordered state, i.e. $m^\ast>0$, which
represents retrieval of pattern one. Both fixed-points, however, need
not be stable against perturbations in the direction of non-nominated
patterns \cite{LV}.  
We have to define initial conditions  that will generate a pure 
macroscopic state, for which we choose 
\be
p_0(\bsigma)=\prod_i\left\{
\frac{1}{2}[1 \plus m(0)] \delta_{\sigma_i,\xi^1_i} +
\frac{1}{2}[1 \minus m(0)] \delta_{\sigma_i,-\xi^1_i}
\right\}
\label{eq:initialLV}
\ee
This indeed gives
$m_\mu^\ast(0)=m(0)\delta_{\mu,1}$, as it should.

\subsection{Statistics of Finite Size Effects}

The restriction to `pure' macroscopic states simplifies our finite
size analysis considerably. The relevant objects in
the Fokker-Planck equation (\ref{eq:fokkerplanck}) become
\bd
\bK(t)= K(t)~\bR,~~~~~~~~~~
L_{\mu\nu}(t) = \ell_\mu(t)\delta_{\mu\nu},~~~~~~~~~~
\bD(t)=D(t)~ \bI
\ed
with the scalar functions
\be
K(t)= -\tanh\beta[w_1 m^\ast(t)]
~~~~~~~~~~~~~~~
\ell_\mu(t) = 1- \beta w_\mu
\left[1\minus\tanh^2\beta[w_1 m^\ast(t)]\right]
\label{eq:KandLforLV}
\ee
\be
D(t)
=  1- \tanh\beta[w_1 m^\ast(t)]\left\{
e^{-t}m(0)
+\int_0^t\!ds~e^{s-t}
\tanh\beta[w_1 m^\ast(s)]
\right\}
\label{eq:diffusionLV}
\ee
and with the (stationary) vector $\bR$, defined as
\be
R_1=0,~~~~~~~~~~
R_{\mu>1}  = \lim_{N \to \infty} \frac{1}{\sqrt{N}}
 \sum_i \xi^\mu_i  \xi^1_i
\label{eq:K0mu}
\ee
The propagator (\ref{eq:propagator}) now becomes trivial:
\be
G_{\mu\nu}(t)=\delta_{\mu\nu}~e^{-\int_0^t\!ds~\ell_\mu(s)}
\label{eq:propagatorLV}
\ee
This will enable us to calculate the moments of the
distribution ${\cal P}_t(\bq)$ at any time explicitly.
Since
(\ref{eq:initialLV}) describes statistically
independent initial components $\sigma_i$, the initial
distribution ${\cal P}_0(\bq)$ is Gaussian.
>From (\ref{eq:initialLV}) it follows that
$\bra \bq \ket_0 = m(0)\bR$ and
$\bXi(0) =[1\minus m^2(0)]\bI$.
The moments at any time $t\geq 0$ then follow from
(\ref{eq:qsolution},\ref{eq:qqsolution}):
\be
\bra q_\mu\ket_t=R_\mu\left\{
m(0) e^{-\int_0^t\!ds~\ell_\mu(s)}
+\int_0^t\!ds~e^{-\int_s^t\!du~\ell_\mu(u)}\tanh\beta[w_1 m^\ast(s)]
\right\}
\label{eq:averagesLV}
\ee
\be
\Xi_{\mu\nu}(t)=\delta_{\mu\nu}\Xi_{\mu}(t),~~~~~~~~
\Xi_\mu(t)=
[1\minus m^2(0)]e^{-2\int_0^t\!ds~\ell_\mu(s)}
+2\int_0^t\!ds~e^{-2\int_s^t\!ds~\ell_\mu(s)}D(s)
\label{eq:correlationsLV}
\ee
This gives the full distribution
\be
{\cal P}_t(\bq)= \left[\prod_\mu[2\pi \Xi_\mu(t)]\right]^{-\frac{1}{2}}
e^{ - \frac{1}{2}\sum_\mu[q_\mu - \bra q_\mu\ket_t]^2/\Xi_\mu(t)}
\label{eq:qdistributionLV}
\ee
which describes uncoupled fluctuations, together with `frozen' finite
size corrections to
the overlaps corresponding to uncondensed patterns.
The above results obviously break down when the propagator develops
runaway solutions, which is likely to happen at phase transitions.

\subsection{Near the Ordering Transition}

We will inspect the behaviour of the finite size effects in stationary
states close to the phase transition separating the paramagnetic from the ordered
state. For stationary states, where
$m^\ast(t)=m^\ast$ for all $t\geq 0$ (with $m^\ast$ given by the solution of
$m^\ast=\tanh\beta[w_1 m^\ast]$), we have in the asymptotic region
(i.e. for $t\to\infty$)  $K(t)=K$,
$\ell_\mu(t)=\ell_\mu$ and
$D(t)=D$, with
\bd
K= -m^\ast
~~~~~~~~~~~~~
\ell_\mu = 1- \beta w_\mu\left[1\minus (m^\ast)^2\right]
~~~~~~~~~~~~~
D=  1- (m^\ast)^2
\ed
This gives
\bd
\bra q_\mu\ket_\infty=\frac{m^\ast R_\mu}{1- \beta w_\mu[1\minus (m^\ast)^2]}
~~~~~~~~~~~~~
\Xi_\mu(\infty)=\frac{1\minus (m^\ast)^2}{1- \beta w_\mu[1\minus (m^\ast)^2]}
\ed
In the paramagnetic state, i.e. $m^\ast=0$, we thus find
\bd
\bra q_\mu\ket_\infty=0,~~~~~~~~
\Xi_\mu(\infty)=\frac{T}{T- w_\mu}
\ed
The fluctuations diverge for $T\downarrow T_c=\max_{\nu}w_\nu$, which
indeed marks the temperature where the paramagnetic state
destabilizes, in favour of a non-trivial pure state.
In a non-trivial pure  state, i.e. for $m^\ast>0$, we find that
both the asymptotic average $\bra q_\mu\ket_\infty$ and
the asymptotic variance $\Xi_\mu(\infty)$ for a fluctuation direction
$\mu$ diverge when $\ell_\mu\to 0$. This again makes sense: the
condition $\ell_\mu>0$ for all $\mu>1$ is the condition for the pure state
corresponding to pattern one to be macroscopically stable,
with $\ell_\mu=0$
signaling destabilization of this pure state in favour of an
alternative pure state $\mu>1$ (see \cite{LV}).
Expansion of the macroscopic fixed-point equation close to $T_c$,
for that particular pure state which is the first to
order as the temperature is lowered (i.e. we now assume
$\max_{\nu}w_\nu=w_1$ and $T_c=w_1$),  gives
\bd
0<\kappa = \frac{T_c-T}{T_c}\ll 1:~~~~~~~~\beta w_1=1+\kappa+\order(\kappa^2),
~~~~~~~
m^\ast=\sqrt{3\kappa}+\order(\kappa)
\ed
This allows us to expand the finite size terms in powers of the
rescaled distance $\kappa$ from the critical temperature:
\bd
~~~~~~~
\bra q_1\ket_\infty=0~~~~~~~~~~~~~~~~~~~~
\bra q_{\mu>1}\ket_\infty=
\frac{\sqrt{3\kappa} R_\mu w_1}{w_1-
w_\mu}+\order(\kappa)
\ed
\bd
\Xi_1(\infty)=\frac{1}{2\kappa}+\order(\kappa^0)
~~~~~~~~~~~~~
\Xi_{\mu>1}(\infty)=\frac{w_1}{w_1- w_\mu}
\ed
The approach of the transition $T\uparrow T_c$ is signalled by
diverging fluctuations, as it should.

\begin{figure}[t]
\centering
\vspace*{127mm}
\hbox to
\hsize{\hspace*{-15mm}\includegraphics{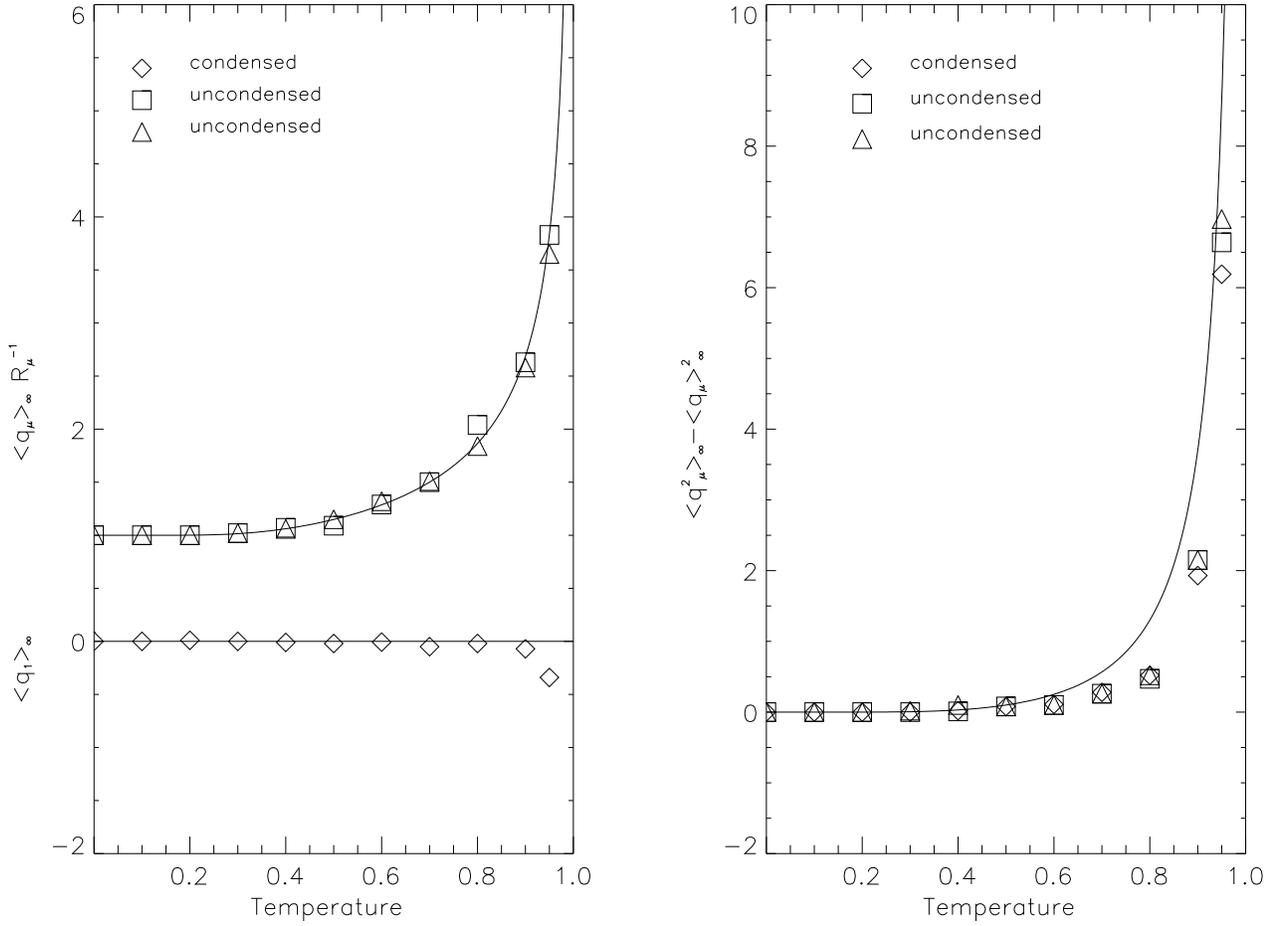}\hspace*{15mm}}
\vspace*{-3mm}
\caption{Comparison between the theoretical predictions for the
moments of the finite-size contributions to the pattern overlaps and numerical
simulations, in stationary states.
Left graph: normalised averages $\bra q_1\ket$ and $\bra
q_\mu\ket_\infty R_\mu^{-1}$ ($\mu>1$). 
Right graph: variances $\bra q_\mu^2\ket_\infty-\bra
q_\mu\ket_\infty^2$. Solid lines: theoretical predictions. Markers:
simulation results, for $N=10 000$ for $T \leq .6$ and $N=50 000
 $ for $T > .6$  }
\label{fig:LVTc}
\end{figure}
Finally we test some of the above predictions against numerical
simulations. The simplest model of our (symmetric) class is the
one where all embedding strengths are equal: $w_\mu=1$  for all $\mu$.
Here we know that at for $T>1$ the system will be paramagnetic,
whereas at $T=1$ a second order thermodynamic transition occurs to a pure
low-temperature state.
If we denote the non-negative
solution of the macroscopic fixed-point equation $m^\ast=\tanh\beta[m^\ast]$
by $m(T)$, with $m(T>1)=0$, we arrive at the following predictions:
\bd
\begin{array}{llllll}
\mu~{\rm condensed}: && \bra q_\mu\ket_\infty=0 &&&
\bra q^2_\mu\ket_\infty -\bra q_\mu\ket_\infty^2=
\frac{T[1- m^2(T)]}{T- 1+ m^2(T)}\\[2mm]
\mu~{\rm uncondensed}: &&
\bra q_\mu\ket_\infty R_\mu^{-1}=
\frac{T m(T)}{T- 1+ m^2(T)}
&&& \bra q^2_\mu\ket_\infty -\bra q_\mu\ket_\infty^2=
\frac{T[1- m^2(T)]}{T- 1+ m^2(T)}
\end{array}
\ed
In figure \ref{fig:LVTc}
we show these predicted equilibrium
moments as functions of temperature, together with results from
numerical simulations carried out for $N=10 000$ for $T \leq .6$
and $N=50 000$ for $T > .6$. The agreement between
theory and experiment is satisfactory. 

\clearpage
\subsection{Zero Temperature}

For $T\to 0$ (or $\beta\to\infty$) the mean-field equation for
the amplitude of pure states
reduces to $\frac{d}{dt}m(t)=1 -m(t)$, given our convention
$m(0)>0$, with solution
\be
m(t)=m(0)e^{-t}+1-e^{-t}
\label{eq:zeroTmacro}
\ee
Since $m(t)>0$ for any time (i.e. we will always be away from the
$T=0$ discontinuities at $m=0$), we may deal with the non-trivial terms
in our problem by using
\bd
\lim_{\beta\to\infty}
\beta [1-\tanh^2\beta[w_1 m(t)]]=w_1^{-1}\lim_{m\to m(t)}
\lim_{\beta\to\infty}\frac{d}{dm}\tanh[\beta w_1 m]
=2w_1^{-1}\lim_{m\to m(t)}\delta(m) = 0
\ed
With this identity we obtain for $T=0$:
\bd
K(t)= -1~~~~~~~~~~\ell_\mu(t) = 1~~~~~~~~~~D(t)=e^{-t}[1-m(0)]
\ed
which, in turn, gives the simple propagator $\bG(t)=\bI~e^{-t}$
and the following moments for the finite size corrections:
\bd
\begin{array}{lllll}
\mu~{\rm condensed}: && \bra q_\mu\ket_t=0 &&
\bra q^2_\mu\ket_t -\bra q_\mu\ket_t^2=
[1\minus m(0)]e^{-t}\left\{2-[1\minus m(0)]e^{-t}
\right\}
\\[2mm]
\mu~{\rm uncondensed}: &&
\bra q_\mu\ket_t R_\mu^{-1}=m(t)
&& \bra q^2_\mu\ket_t -\bra q_\mu\ket_t^2=
[1\minus m(0)]e^{-t}\left\{2-[1\minus m(0)]e^{-t}
\right\}
\end{array}
\ed
\begin{figure}[t]
\centering
\vspace*{127mm}
\hbox to
\hsize{\hspace*{-15mm}\includegraphics{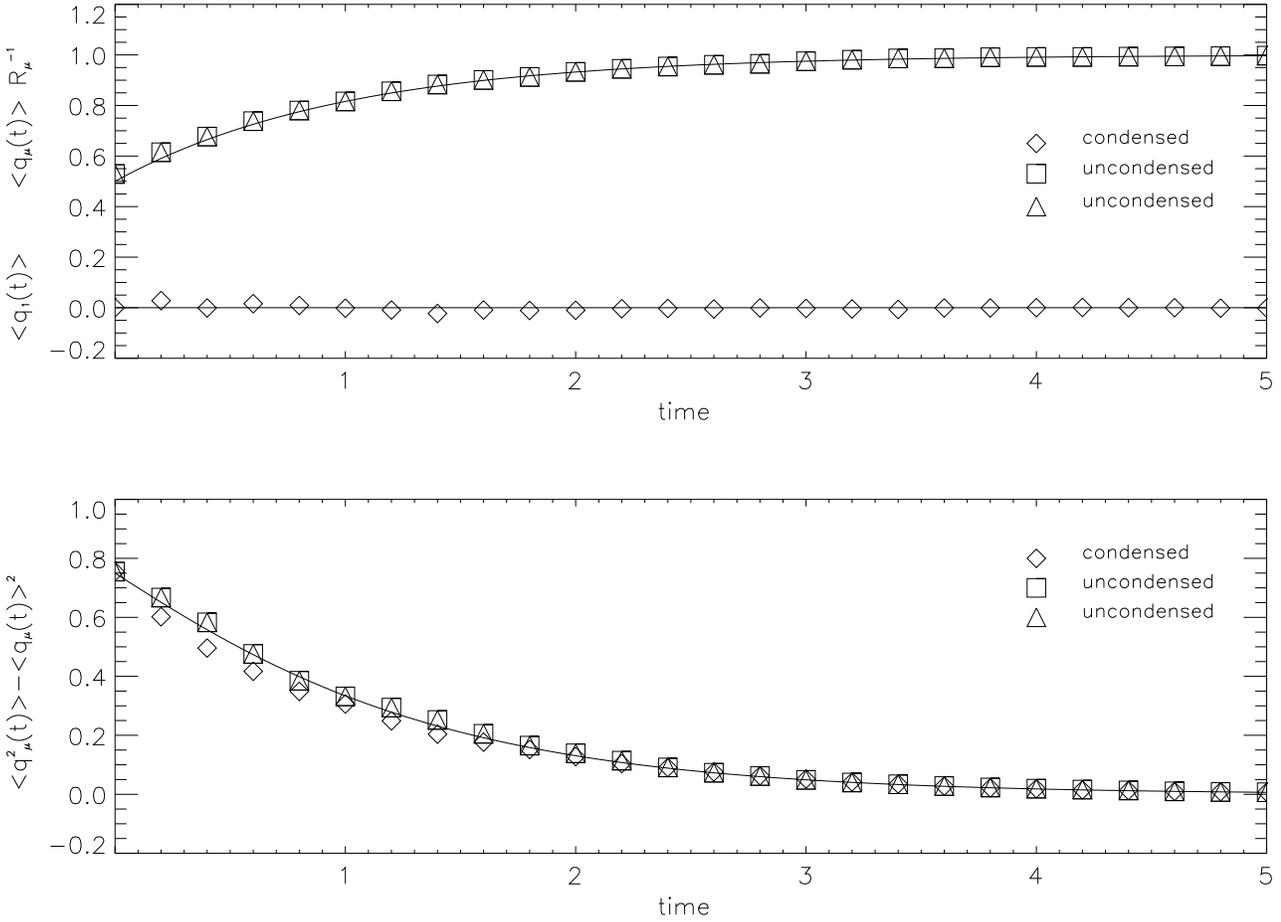}\hspace*{15mm}}
\vspace*{-3mm}
\caption{Comparison between the theoretical predictions for the
moments of the finite-size contributions to the pattern overlaps and
numerical 
simulations, at zero temperature, as functions of time.
Upper graph: normalised averages $\bra q_1\ket$ and  $\bra q_\mu\ket
R_\mu^{-1}$ ($\mu>1$). 
Lower graph: variances $\bra q_\mu^2\ket-\bra
q_\mu\ket^2$. Solid lines: theoretical predictions. Markers:
simulation results, for $N=5 000$.}
\label{fig:LVTzero}
\end{figure}
with $m(t)$ given in (\ref{eq:zeroTmacro}).
The `frozen' correction to the mean-field laws, i.e. the term $\bra
q_\mu\ket_t$,
increases in absolute strength as time progresses. The
fluctuations, which at $T=0$ have their origin purely in the randomness
of the order of the single-spin updates, decrease to zero exponentially.
In figure \ref{fig:LVTzero} we show these predicted zero temperature
moments as functions of time, together with results from
numerical simulations carried out for $N=5 000$. Again the agreement
between theory and experiment is quite satisfactory. 

\section{Application to Non-Equilibrium Models}

In this section  we will apply our theory to non-symmetric systems,
i.e. $A_{\mu\nu}\neq A_{\nu\mu}$, for which detailed balance does not
hold. We will restrict ourselves to the case $p=2$ and
$W(\theta)=\delta(\theta)$ for simplicity;
extensions to larger values of $p$ and/or non-zero external fields are straightforward and are not
likely to generate new physics.
We define an initial microscopic distribution with statistically
independent spins, in order to guarantee a Gaussian shape for ${\cal
P}_0(\bq)$, given by
\be
p_0(\bsigma)=\prod_i\left\{\frac{1}{2}[1\minus m_1(0)\minus
m_2(0)]
+m_1(0)\delta_{\sigma_i,\xi_i^1}+m_2(0)\delta_{\sigma_i,\xi_i^2}\right\}
\label{eq:Psigma2}
\ee
Obviously, we have to restrict ourselves to the physical region, defined by
the two conditions $|m_1(0)+m_2(0)|\leq 1$ and $|m_1(0)-m_2(0)|\leq
1$.
Our definition generates the required initial macroscopic
observables, $\lim_{N\to\infty}\bra m_\mu(\bsigma)\ket_0=m_\mu(0)$
($\mu=1,2$),
and gives the following initial moments for the finite
size variables:
\be
\left(\!\begin{array}{c}
\bra q_1\ket_0 \\ \bra q_2\ket_0 \end{array}\!\right)=
R\left(\!\begin{array}{c} m_2(0) \\ m_1(0) \end{array}\!\right)
~~~~~~~~~~~
\bXi(0)=\left(\!\begin{array}{cc}1-\bm^2(0)
& -2 m_1(0)m_2(0)\\ -2 m_1(0)m_2(0) &
1-\bm^2(0) \end{array}\!\right)
\label{eq:inmitialmomentsnobalance}
\ee
with $\bm(0)=(m_1(0),m_2(0))$ and with
$R=\lim_{N\to\infty}\frac{1}{\sqrt{N}}\sum_i\xi_i^1\xi_i^2$.

\subsection{Non-Equilibrium Stationary States}

In this subsection we will study the class of networks where the
matrix $\bA$ has the form
\bd
\bA=\left(\!\!\begin{array}{cc}1 & \epsilon \\ 0 &
1\end{array}\!\!\right)
\ed
with $\epsilon\geq 0$.
These systems obey detailed balance only for $\epsilon=0$. The
mean-field equations, describing the overlap evolution
in the $N \to \infty$ limit, are given by
\be
\frac{d}{dt}\bm^\ast = -\bm^\ast
+\frac{1}{2}\tanh\beta[m_1^\ast\plus (\epsilon\plus 1)m_2^\ast]
\left(\!\!\begin{array}{c}1\\ 1\end{array}\!\!\right)
+\frac{1}{2}\tanh\beta[m_1^\ast\plus (\epsilon\minus 1)m_2^\ast]
\left(\!\!\begin{array}{c}1\\ -1\end{array}\!\!\right)
\label{eq:meanfieldnobal}
\ee
In the low temperature regime  these equations have two types
of fixed-points. Firstly, for $T<1$ one finds the two non-trivial
fixed-points 
$\bm^\ast =\pm (m^\ast, 0)$, related only to pattern one,
where $m^\ast$ is the positive solution of  $m^\ast= \tanh [ \beta
m^\ast]$.
The second set of fixed-points are related to pattern two.
For $\epsilon = 0$, these are simply the
 pure states $\bm^\ast =\pm (0, m^\ast)$.
For $\epsilon\neq 0$, however, they move away from
the $m^\ast_1=0$ axis in the $(m^\ast_1,m^\ast_2)$ plane, towards
the pattern one pure states, until the fixed-points merge pair-wise.
The magnitude of this displacement increases with $\epsilon$; the
rate of increase being low  for small $\epsilon$, but large close
to the point where the fixed-points merge. The value of $\epsilon$
where merger occurs decreases with increasing
temperature.
In all cases the second class of fixed-points is
found to have disappeared for $\epsilon =1$.

We will study the finite size effects for the pure
fixed-point $\bm^\ast=(m^\ast,0)$, and their dependence on the
parameter $\epsilon$ which can be interpreted as measuring the
degree of violation of detailed balance. To assess the macroscopic
stability of this pure state we study the effect of
perturbations:
$\bm^\ast(t)=(m^\ast,0)+(\delta_1(t),\delta_2(t))$, with
$|\delta_1(t)|\ll 1$ and $|\delta_2(t)|\ll 1$. Linearisation of the
mean-field laws gives
\bd
\frac{d}{dt}\left(\!\!\begin{array}{c}\delta_1\\ \delta_2\end{array}\!\!\right)
=\left\{\left[\beta[1-(m^\ast)^2]\minus 1\right]\bI+
\beta\epsilon[1\minus(m^\ast)^2]
\left(\!\!\begin{array}{cc}0 & 1\\ 0 &
0\end{array}\!\!\right)\right\}
\left(\!\!\begin{array}{c}\delta_1 \\ \delta_2\end{array}\!\!\right)
+\ldots
\ed
with the solution
\bd
\left(\!\!\begin{array}{c}\delta_1(t)\\ \delta_2(t)\end{array}\!\!\right)
=e^{\left[\beta[1-(m^\ast)^2]- 1\right]t}
\left(\!\!\begin{array}{c}\delta_1(0)+\beta\epsilon t[1-(m^\ast)^2]\delta_2(0)
\\ \delta_2(0)\end{array}\!\!\right)
\ed
Therefore the pure states $\bm^\ast=(m^\ast,0)$ are (globally)
stable if and
only if $\beta [1\minus (m^\ast)^2]-1<0$. This condition is met by
the solution of $m^\ast=\tanh[\beta m^\ast]$ as soon as it is
non-zero, i.e. for all $T<1$.

For finite $N$ the mean-field picture will be modified by finite size
effects.
In the pure fixed-point $\bm^\ast=(m^\ast,0)$, obtained following a pure initialisation
$\bm^\ast(0)=(m^\ast(0),0)$, the stationary
Ornstein-Uhlenbeck process is characterised by
\bd
\bL=\bI- \beta [1\minus (m^\ast)^2]\bA
~~~~~~~~~
\bD= \left[1- (m^\ast)^2\right]~\bI
~~~~~~~~~
\bK=
- m^\ast R \left(\!\!\begin{array}
{c}0\\ 1\end{array}\!\!\right)
\ed
We work out the propagator $\bG(t)$ (\ref{eq:propagator}) by
splitting the matrix $\bL$ into two commuting parts, such that
$\bG(t)=\exp[-t\bL]$ factorises into two separate matrix
exponentiations:
\bd
\bG(t)
=\exp\left\{-t[1\minus\beta [1\minus(m^\ast)^2]]\bI
+\epsilon\beta t[1\minus (m^\ast)^2]
\left(\!\!\begin{array}{cc}0 & 1 \\ 0 &
0\end{array}\!\!\right)\right\}
\ed
\bd
=e^{-t[1\minus\beta [1\minus(m^\ast)^2]]}
\left[\bI+\epsilon\beta t [1\minus (m^\ast)^2]
\left(\!\!\begin{array}{cc}0 & 1 \\ 0 &
0\end{array}\!\!\right)\right]
\ed
The condition for the propagator to be well-behaved
is identical to the condition for the fixed-point under
consideration to be macroscopically stable:
$\beta [1\minus (m^\ast)^2]<1$.
By working
out (\ref{eq:qsolution},\ref{eq:qqsolution}) we arrive at
the moments of the
distribution ${\cal P}_t(\bq)$:
\bd
\bra \bq\ket_t=m^\ast(0)R e^{-t\bL}\left(\!\begin{array}{c} 0 \\ 1
\end{array}\!\right)
+ m^\ast R \bL^{-1}[\bI\minus e^{-t\bL}]\left(\!\!\begin{array}
{c}0\\ 1\end{array}\!\!\right)
\ed
\bd
\bXi(t)=e^{-t\bL}
\left(\!\begin{array}{cc}1-\bm^2(0)
& -2 m_1(0)m_2(0)\\ -2 m_1(0)m_2(0) &
1-\bm^2(0) \end{array}\!\right)
e^{-t\bL^\dag}+2\left[1\minus (m^\ast)^2\right]
\int_0^t\!ds~e^{(s-t)\bL}
e^{(s-t)\bL^\dag}
\ed
The limits $t\to\infty$ are given by:
\bd
\bra \bq\ket_\infty=
m^\ast R\left\{
[1\minus\beta [1\minus(m^\ast)^2]]\bI
-\epsilon\beta [1\minus (m^\ast)^2]
\left(\!\!\begin{array}{cc}0 & 1 \\ 0 &
0\end{array}\!\!\right)\right\}^{-1}
\left(\!\!\begin{array}
{c}0\\ 1\end{array}\!\!\right)
\ed
\be
=\frac{m^\ast R}{[1-\beta [1\minus(m^\ast)^2]]^2}
\left(\!\!\begin{array}
{c}\epsilon\beta [1\minus (m^\ast)^2]\\
1-\beta [1\minus(m^\ast)^2]
\end{array}
\!\!\right)
\label{eq:qnobalance}
\ee
\bd
\bXi(\infty)=
2\left[1\minus (m^\ast)^2\right]\lim_{t\to\infty}
\int_0^t\!ds~
e^{2(s-t)[1\minus\beta [1\minus(m^\ast)^2]]}
\left\{\bI
-\epsilon\beta (s\minus t) [1\minus (m^\ast)^2]
\left(\!\!\begin{array}{cc}0 & 1 \\ 1 &
0\end{array}\!\!\right)
\right.
\ed
\bd
\left.
~~~~~~~~~~~~~~~~~~~~~~~~~~~~~~~~~~~~~~~~~~~~~~~~~~~~~~~~~~~~~~~~~
+\epsilon^2\beta^2 (s\minus t)^2 [1\minus (m^\ast)^2]^2
\left(\!\!\begin{array}{cc}1 & 0 \\ 0 &
0\end{array}\!\!\right)
\right\}
\ed
\be
=
H(T)\left(\!\!\begin{array}{cc}1+\frac{1}{2}\epsilon^2\beta^2 H^2(T)
 & \frac{1}{2}\epsilon\beta H(T) \\[2mm] \frac{1}{2}\epsilon\beta H(T) &
1\end{array}\!\!\right)
\label{eq:qqnobalance}
\ee
in which
\bd
H(T)=
\frac{1\minus (m^\ast)^2}{1-\beta [1\minus(m^\ast)^2]},~~~~~~~~~~
H(0)=0,~~H(1)=\infty
\ed
Note that $H(T>1)=T/(T\minus 1)$.
Apart from inducing a non-zero stationary correction $\bra
q_1\ket_\infty$ to the overlap with pattern one,
violation of detailed balance (i.e. having $\epsilon>0$ rather than $\epsilon=0$)
leads to an increase in the fluctuations of the non-trivial
overlap, and a coupling of the fluctuations in the $q_1$ and $q_2$
directions, which in the case of detailed balance would have been
statistically independent.

\begin{figure}[t]
\centering
\vspace*{127mm}
\hbox to
\hsize{\hspace*{-15mm}\includegraphics{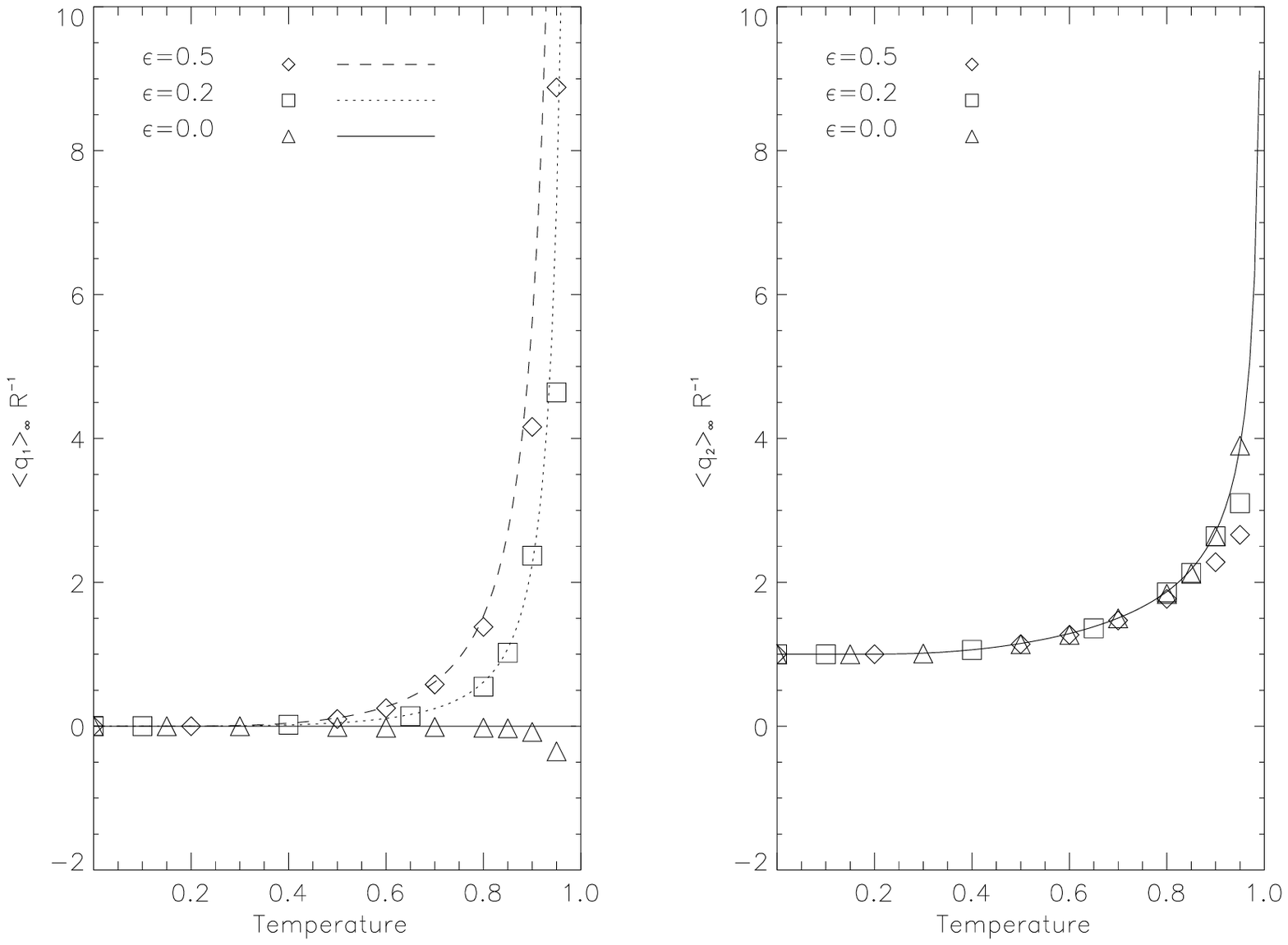}\hspace*{15mm}}
\vspace*{-3mm}
\caption{Comparsion between the theoretical predictions for the first
order moments of the finite-size contributions to the pattern overlaps and numerical
simulations, in stationary states, as functions of temperature and for
$\epsilon\in\{0,~0.2,~0.5\}$. 
Left graph: normalised average $\bra q_{1}\ket_\infty R^{-1}$.
Right graph: normalised average $\bra q_{2}\ket_\infty R^{-1}$.
Solid lines: theoretical predictions. Markers: 
simulation results, for $N=50 000$.} 
\label{fig:nobalance}
\end{figure}
We can appreciate most clearly
the effects of the correlations in the fluctuations by examining the
curl of the probability current $\bJ(\bq)$ in $\bq$-space.
In doing so we can use the
stationarity condition (\ref{eq:stationarityofP})
 for the correlation matrix, i.e.
$\frac{1}{2}[\bL\bXi +(\bL\bXi)^\dag]=\bD$ (which allows
us to put $\bx\cdot[\bD-\bL\bXi]\bx=0$ for each $\bx\in\Re^2$), as well
as the symmetry of both the correlation matrix $\bXi$ and its inverse.
For stationary states the probability current (\ref{eq:current}),
which must be divergence-free,
reduces to
\bd
\bJ(\bq)={\cal P}(\bq)~
\left[\bD\bXi^{-1}- \bL\right](\bq- \bra \bq\ket_\infty)
\ed
Its curl is found to be
\bd
\bnabla\! \times\! \bJ(\bq)=
{\cal P}(\bq)\left\{ \bnabla\! \times\! \left[\bD
\bXi^{-1}\minus \bL\right](\bq\minus  \bra \bq\ket_\infty)
-(\bq\minus \bra\bq\ket_\infty)\!\cdot\!(\bXi^{-1})^\dag
\left[\bD\minus \bL\bXi\right]\bXi^{-1}(\bq\minus \bra
\bq\ket_\infty)
\right\}
\ed
\bd
=
{\cal P}(\bq)~\bnabla \!\times\! \left\{
[\bD\bXi^{-1}\minus \bL]
(\bq\minus \bra \bq\ket_\infty) \right\}
\ed
\bd
=-\epsilon\beta [1\minus(m^\ast)^2]~{\cal P}(\bq)
\ed
Violation of detailed balance, due to the asymmetry of the matrix
$\bA$ for $\epsilon\neq 0$,
produces a stationary rotational current in the space of the finite size
variables $\bq$.
The magnitude of this
current is proportional to
the magnitude of the parameter $\epsilon$.
For $T<1$ the prefactor $\beta[1\minus (m^\ast)^2]$ is a monotonically
increasing function of temperature, starting at zero for $T=0$ and
approaching one for $T\to 1$.
The rotational current
persists above the critical temperature, where we find
$\bnabla\!\times\!\bJ(\bq)=-\epsilon\beta~{\cal P}(\bq)$.

Finally we test the predictions
(\ref{eq:qnobalance},\ref{eq:qqnobalance}) for the
moments of ${\cal P}_\infty(\bq)$
 against numerical
simulations.
In figure \ref{fig:nobalance}
we show these predicted equilibrium
moments as functions of temperature, together with results from
numerical simulations carried out for $N=50 000$ and 
$\epsilon\in\{0,0.2,0.5\}$.
The present non-equilibrium 
model is found to require larger system sizes for our
fluctuation theory to hold (i.e. neglected higher orders in $N^{-\frac{1}{2}}$
are more prominent) than the equilibrium models studied in earlier
sections. In addition the time required for transient effects to have
died out is longer. This explains why the agreement between theory and
experiment, as observed in figure \ref{fig:nobalance}, although still 
reasonable, is less than that in observed in our previous experiments.

\clearpage
\subsection{Escape Times Controlled by System Size}

As a final application of our finite size theory we turn to a model
which at $T=0$ is exactly critical, in the sense that the mean-field
flow is such that the asymptotic value $\bm^\ast(\infty)$
of the overlap vector is exactly
{\em on} a regional boundary in $\bm$-space which separates
qualitatively different  macroscopic
flow domains:
\bd
\bA=\left(\!\!\begin{array}{cc}1 & -1 \\ 1 & 1\end{array}\!\!\right)
\ed
As a result one finds in this particular system
that the relevant escape times, which dictate
whether and when the state vector can leave  a given domain, are
controlled entirely by the size $N$ of the system.

At non-zero noise levels $0<T<1$ the solutions of the mean-field
equations for $\bm^\ast$, describing
the overlap evolution in the $N\to\infty$ limit, show evolution into
a stable limit-cycle \cite{separable1}.
For $T=0$ the mean-field equations reduce to
\bd
\frac{d}{dt}\bm^\ast=
\frac{1}{2}\sgn[m^\ast_1]\left(\!\!\begin{array}{c}1\\1\end{array}\!\!\right)
+\frac{1}{2}\sgn[m^\ast_2]\left(\!\!\begin{array}{c}-1\\1\end{array}\!\!\right)
-\bm^\ast
\ed
giving
\be
\begin{array}{llll}
{\rm region~I}, &
m_1^\ast(0)>0,~m_2^\ast(0)>0: && \bm^\ast(t)=\bm^\ast(0)e^{-t}+
(0,1)[1-e^{-t}]\\[2mm]
{\rm region~II}, &
m_1^\ast(0)<0,~m_2^\ast(0)>0: && \bm^\ast(t)=\bm^\ast(0)e^{-t}+
(\minus 1,0)[1-e^{-t}]\\[2mm]
{\rm region~III}, &
m_1^\ast(0)<0,~m_2^\ast(0)<0: && \bm^\ast(t)=\bm^\ast(0)e^{-t}+
(0,\minus 1)[1-e^{-t}]\\[2mm]
{\rm region~IV}, &
m_1^\ast(0)>0,~m_2^\ast(0)<0: && \bm^\ast(t)=\bm^\ast(0)e^{-t}+
(1,0)[1-e^{-t}]\\[2mm]
\end{array}
\ee
There are four qualitatively different macroscopic flow
regions,
separated by the two lines $m^\ast_1=0$ and $m_2^\ast=0$.
In all four cases the macroscopic flow is
directed towards a state which is exactly {\em at} the regional
boundary, such that the asymptotics of the system (i.e. whether or not
the state vector will escape to another region) will be decided purely
by the finite size effects.

Due to the overall symmetry of our
models with respect to the transformation
$\forall\bsigma:~p_t(\bsigma)\to p_t(-\bsigma)$ (at least: in the absence of external
fields), the properties of regions I and III and of regions II and
IV are pair-wise identical. Furthermore,  the properties of region II
follow from those of region I via the transformation $\bxi^1\to
-\bxi^1$, and the properties of region IV
follow from those of region III via the transformation $\bxi^1\to
-\bxi^1$. This implies that without loss of generality
we can restrict our quantitative
analysis to region I.
We choose $T=0$ and the initial state  $\bm^\ast(0)=(m^\ast(0),0)$,
with $0<m^\ast(0)<1$ (i.e. in region I). The relevant quantities
in the Fokker-Planck equation (\ref{eq:fokkerplanck}) are then
given by
\bd
\bK(t)= -\left(\!\!\begin{array}{c}R\\0\end{array}\!\!\right)
~~~~~~~~~~~~~
\bL(t) = \bI
~~~~~~~~~~~~~
\bD(t)=e^{-t}\left(\!\!\begin{array}{cc}1 & -m^\ast(0) \\-m^\ast(0) & 1
\end{array}\!\!\right)
\ed
We find the simple propagator $\bG(t)=e^{-t}\bI$.
Note that working out the relevant moments
(\ref{eq:qsolution},\ref{eq:qqsolution}) of ${\cal P}_t(\bq)$ for
the present model, following the initial conditions
(\ref{eq:inmitialmomentsnobalance}), gives
\bd
\bra \bq\ket_t=e^{-t}
Rm^\ast(0)\left(\!\begin{array}{c} 0 \\ 1\end{array}\!\right)
+[1-e^{-t}]R
\left(\!\!\begin{array}{c}1\\0\end{array}\!\!\right)
\ed
\bd
\bXi(t)=e^{-2t}\left(\!\begin{array}{cc}1-(m^\ast(0))^2
& 0\\ 0 & 1-(m^\ast(0))^2 \end{array}\!\right)
+2e^{-t}[1-e^{-t}]\left(\!\!\begin{array}{cc}1 & -m^\ast(0) \\-m^\ast(0) & 1
\end{array}\!\!\right)
\ed
We want to
calculate the probability that at time $t$ the system will have escaped from
region I to region II. To this aim we first define
\be
\Pi_t (M) =
\mbox{Prob} [m_1(\bsigma(t))< M ]=
\sum_{\bsigma}p_t(\bsigma)\theta[M-m_1(\bsigma)]
\label{eq:generalpi}
\ee
Note that in both regions I and II we have $m_2(\bsigma)>0$ (in
fact the escape process I$\to$II happens close to
$m_2(\bsigma)=1$).
The  time derivative of $\Pi_t(M)$ follows from
(\ref{eq:ddtaverage}), which for $T=0$ reduces to
\bd
\frac{d}{dt}\Pi_t(M)=\bra \left\{M-
\frac{1}{N}\sum_i
\sgn\left[M\minus m_2(\bsigma)+\xi_i^1\xi_i^2[M\plus m_2(\bsigma)]\right]
\right\}
\delta[M-m_1(\bsigma)]\ket_t
+\order(N^{-1})
\ed
\bd
=\bra\left\{ M +\theta[\minus M]
-\frac{R}{\sqrt{N}}\theta[M]
\right\}\delta[M-m_1(\bsigma)]\ket_t
+\order(N^{-1})
\ed
with the usual definition
$R=\frac{1}{\sqrt{N}}\sum_i\xi_i^1\xi_i^2$.
It follows that $\frac{d}{dt}\Pi_t(M)$ is discontinuous at $M=0$:
\begin{eqnarray*}
\lim_{M \to 0_+} \frac{d}{dt} \Pi_t (M) &\!=\!& -
\frac{R}{\sqrt{N}}
\bra \delta [m_1 (\sigma)] \ket_t +\order(N^{-1})\\
\lim_{M \to 0_-} \frac{d}{dt} \Pi_t (M) &\!=\!&
\bra \delta [m_1 (\sigma)] \ket_t +\order(N^{-1})
\end{eqnarray*}
The escape process requires times sufficiently large to allow
finite size effects to come into play, i.e.
$e^{-t}=\order(N^{-\frac{1}{2}})$. We are thus led to the
introduction of the new time variable\footnote{
Here there could be a potential conflict with the assumptions of
the theory. However, inspection shows that our derivation of the
Fokker-Planck equation as the correct description of the leading
order finite size effects
requires $\lim_{N\to\infty}t/\sqrt{N}=0$, which means that the theory
still applies if $t=\order(\log(N))$.}: $\gamma=e^t/\sqrt{N}$. 
For such times the average $\bra \delta[m_1 (\sigma)]\ket_t$
can be written as
\bd
\bra \delta[m_1 (\sigma)]\ket_t=
\int\!d\bq~{\cal P}_t(\bq)\delta[m^\ast(0)e^{-t}
+\frac{1}{\sqrt{N}}q_1+\order(N^{-1})]
\ed
\bd
=\frac{\sqrt{N}}{\sqrt{2\pi\Xi_{11}}}~
e^{-\frac{1}{2}\left[\bra q_1\ket_t+\frac{m^\ast(0)}{\gamma}
+\order(N^{-\frac{1}{2}})\right]^2/\Xi_{11}(t)}
\ed
In terms of the new time variable we write $\Pi_t(M)=P_\gamma(M)$;
using $\frac{d}{dt}=\gamma\frac{d}{d\gamma}$ we then arrive at
\begin{eqnarray*}
\lim_{M \to 0_+} \frac{d}{d\gamma} P_\gamma (M) &\!=\!& -
\frac{R}{\gamma\sqrt{2\pi\Xi_{11}}}~
e^{-\frac{1}{2}\left[\bra q_1\ket_t+\frac{m^\ast(0)}{\gamma}
+\order(N^{-\frac{1}{2}})\right]^2/\Xi_{11}(t)}
 +\order(N^{-1})\\
\lim_{M \to 0_-} \frac{d}{d\gamma} P_\gamma (M)
&\!=\!&
\frac{\sqrt{N}}{\gamma\sqrt{2\pi\Xi_{11}}}~
e^{-\frac{1}{2}\left[\bra q_1\ket_t+\frac{m^\ast(0)}{\gamma}
+\order(N^{-\frac{1}{2}})\right]^2/\Xi_{11}(t)}
 +\order(N^{0})
\end{eqnarray*}
Note that on the $\gamma$ time-scales 
$\bra q_1\ket_t=R+\order(N^{-\frac{1}{2}})$ and 
$\Xi_{11}(t)=\order(N^{-\frac{1}{2}})$. Consequently
\begin{eqnarray*}
\frac{d}{d\gamma} P_\gamma (0_+) &\!=\!& -
\frac{R}{\gamma}~\delta\left[R+\frac{m^\ast(0)}{\gamma}\right]
 +\order(N^{-\frac{1}{2}})\\
\frac{d}{d\gamma} P_\gamma (0_-) &\!=\!&
\frac{\sqrt{N}}{\gamma}~\delta\left[R+\frac{m^\ast(0)}{\gamma}\right]
+\order(N^{0})
\end{eqnarray*}
\begin{figure}[t]
\centering
\vspace*{122mm}
\hbox to
\hsize{\hspace*{-15mm}\includegraphics{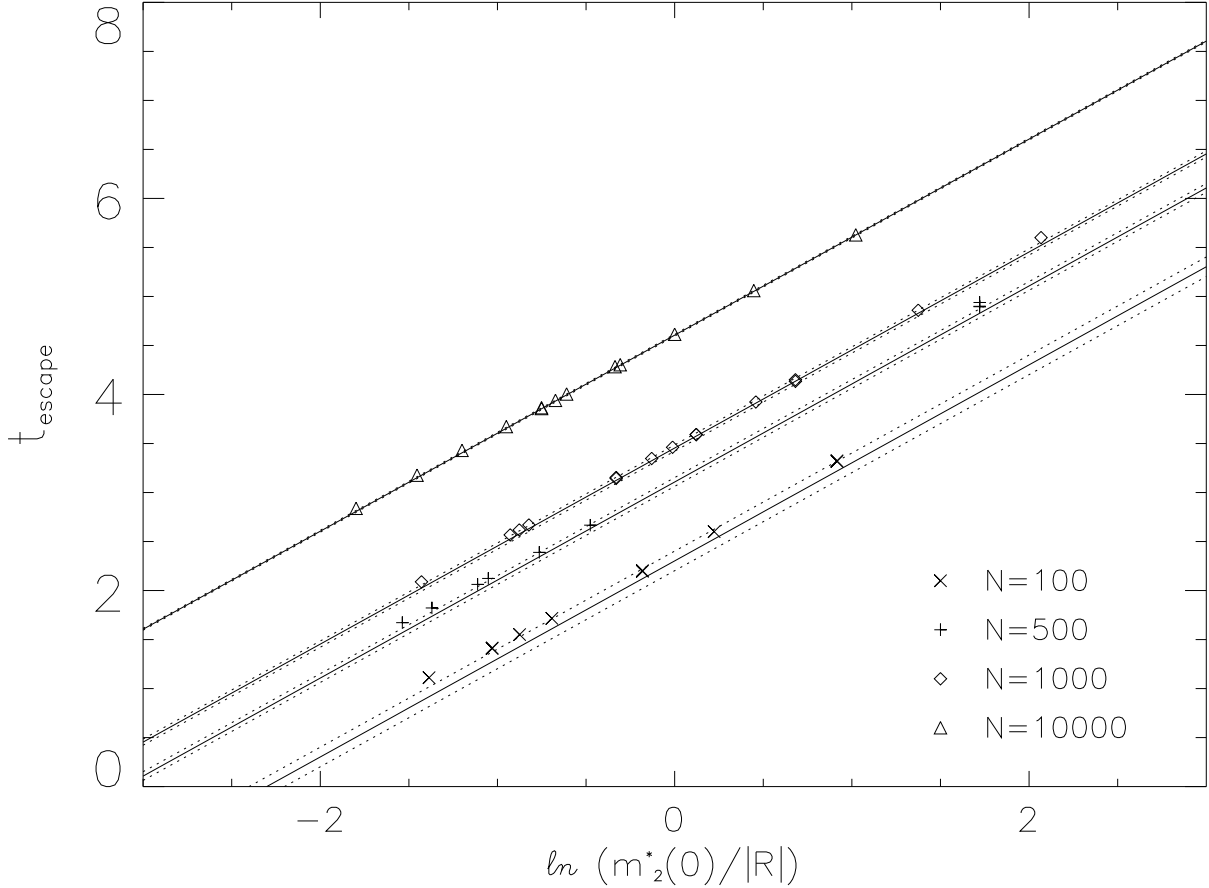}\hspace*{15mm}}
\vspace*{-3mm}
\caption{Comparison between the theoretical prediction
(\ref{eq:escapetime})
for the 
escape time from the initial region and numerical simulations.
Solid lines: leading two orders according to the theory. Dashed lines:
indication of the potential magnitude of subsequent (neglected)
orders $\order(N^{-\frac{1}{2}})$. Markers: simulation results.}
\label{fig:escapetime}
\end{figure}
Since $P_0(0_\pm)=0$ (for we start with $m_1^\ast(0)>0$) we are
led to the following predictions.
If $R> 0$ the state vector never escapes region I; if $R<0$,
on the other hand, the state vector will ultimately escape. Once past the
regional boundary, the state vector cannot return due to the boost
described by $P_\gamma(0_-)$. Integration over $\gamma$
gives the explicit form
\bd
P_\gamma (0^+) =-R
\int_{m^\ast(0)/\gamma}^\infty \!\frac{dz}{z}~\delta\left[R+z\right]
 +\order(N^{-\frac{1}{2}})
 \ed
 \bd
=\theta[-R] \theta\left[\gamma -\frac{m^\ast(0)}{|R|}\right]
+\order(N^{-\frac{1}{2}})
\ed
So, provided $R<0$, the state vector leaves region I precisely
when $\gamma = m^\ast(0)/|R|$. Translation back into the original
time variable gives the following escape time:
\be
t_{\rm esc} = \frac{1}{2} \log N +
\log\left[\frac{m^\ast(0)}{|R|}\right]
+ \order \left( \frac{1}{\sqrt{N}} \right)
\label{eq:escapetime}
\ee
If we use the simple transformations that relate the properties of
the four regions, we arrive at
the following picture. If $R< 0$ the system will be able to make
the transitions I$\to$II and III$\to$IV, but never the transitions
II$\to$III or IV$\to$I. For $R> 0$, on the other hand,
the transitions II$\to$III and IV$\to$I will be observed, but never the transitions
I$\to$II or III$\to$IV. When $R=0$, the case where the two patterns are
orthogonal in the first two leading orders in $N$,
the escape properties will be controlled by the $\order(N^{-1})$ finite
size effects.

Finally we compare the prediction (\ref{eq:escapetime}) for the
escape time with the results of numerical simulations.
Figure \ref{fig:escapetime} shows  the average escape time as a function of
$\log [m^\ast(0) /|R|]$. The agreement between theory and experiments
is quite satisfactory.

\section{Discussion}

We have performed a systematic study of 
finite size effects in separable recurrent neural
network models
away from saturation. Since our 
approach is based on analysis of the dynamics, our results apply to
models with  detailed balance (i.e. with symmetric synaptic
interactions) and to models without detailed balance (with
non-symmetric synaptic interactions). 
 In leading order in the system size
($N^{-\frac{1}{2}}$) the finite size effects
 turn out to
be governed by a time-dependent Ornstein-Uhlenbeck process, and
their time-dependent probability density can be calculated in
explicit form. The leading order finite size effects are found to come in two distinct
forms: they show up as `frozen' corrections to the mean field laws
(dependent on the details of the correlations between the randomly
drawn stored patterns) and as fluctuations, which have their origin in
thermal noise in the local field alignment as well as the randomness
in the selection of the neuron to be updated. 

We use our theory to work out several specific but characteristic examples,
including symmetric attractor neural network models, used as
associative memories, and non-equilibrium  models (with non-symmetric
interactions). For detailed balance models we quantify within our
fluctuation theory the familiar features of equilibrium statistical
mechanics, e.g. diverging fluctuations near phase transitions and
absent probability currents in the stationary state. 
For non-equilibrium models, in contrast, we find persistent rotational
currents in the stationary state.   
One of our non-equilibrium examples involves the calculation of
escape times which are purely controlled by finite size effects, which
is a nice example  of a problem where the finite size effects are
significantly more than 
simply a correction to the corresponding result for an infinite
system. More extensive applications will  be published in
\cite{thesis}. 
Comparison with extensive numerical simulations confirms the 
theoretical predictions in all cases. 

\subsection*{Acknowledgements}

One of the authors (ACCC) would like to thank the UNAM Institute 
of Physics in Ensenada (now CECIMAT) for their kind hospitality. 
This work was partially supported by grant DGAPA IN100895 (UNAM).

\end{document}